\documentclass[%
 reprint,
superscriptaddress,
 amsmath,amssymb,
 prl,
]{revtex4-1}

\usepackage{amssymb}
\usepackage{amsmath}
\usepackage{graphicx}
\usepackage{epsfig}
\usepackage{rotating}
\usepackage{verbatim}
\usepackage{wrapfig}
\usepackage{float}
\usepackage[export]{adjustbox}
\usepackage{subcaption}
\usepackage[skip=2pt,font=small]{caption}
\usepackage{ftnxtra}
\usepackage{physics}
\usepackage{graphicx}
\usepackage{dcolumn}
\usepackage{bm}
\newcommand{\bld}[1]{\boldsymbol{#1}}
\providecommand\bcdot{\boldsymbol{\cdot}}
\newcommand\bnabla{\boldsymbol{\nabla}}

\usepackage[
total={6.5in,9.75in}, top=0.9 in, left=0.9in, includefoot,
]{geometry}
\begin{document}

\preprint{APS/123-QED}

\title{Viscoelastic pipe flow is linearly unstable}

\author{Piyush Garg}
\affiliation{Engineering Mechanics Unit, Jawaharlal Nehru Centre for Advanced Scientific Research,
Bangalore 560064, India}
\author{Indresh Chaudhary}
 \affiliation{Department of Chemical Engineering, Indian Institute of Technology, Kanpur 208016, India.}
 \author{Mohammad Khalid}
 \affiliation{Department of Chemical Engineering, Indian Institute of Technology, Kanpur 208016, India.}
\author{V Shankar}
 \email{vshankar@iitk.ac.in}
 \affiliation{Department of Chemical Engineering, Indian Institute of Technology, Kanpur 208016, India.}
\author{Ganesh Subramanian}%
 \email{sganesh@jncasr.ac.in}
\affiliation{Engineering Mechanics Unit, Jawaharlal Nehru Centre for Advanced Scientific Research,
Bangalore 560064, India}


\begin{abstract}

Newtonian pipe flow is known to be linearly stable at all Reynolds numbers. We report, for the first time, a linear instability of pressure-driven pipe flow of a viscoelastic fluid, obeying the Oldroyd-B constitutive equation commonly used to model dilute polymer solutions. The instability is shown to exist at Reynolds numbers significantly lower than those at which transition to turbulence is typically observed for Newtonian pipe flow. Our results qualitatively explain experimental observations of transition to turbulence in pipe flow of dilute polymer solutions at flow rates where Newtonian turbulence is absent. The instability discussed here should form the first stage in a hitherto unexplored dynamical pathway to turbulence in polymer solutions. An analogous instability exists for plane Poiseuille flow.

\end{abstract}

\maketitle



Since the discovery by Toms that the addition of small amounts of a high molecular weight polymer to a Newtonian fluid significantly reduces the pressure drop in turbulent pipe flow\cite{toms1977,virk75}, turbulent flows of dilute polymer solutions have been widely studied for both their fundamental and industrial importance \cite{virk75,LumleyAFM1969,toms1977,bermanAFM1978,mung08,alaskanpipeline}. Understanding the transition to turbulence in shearing flows of viscoelastic fluids, including dilute polymer solutions, is thus crucial \cite{graham2014,moro07}. A central question underlying this field of study is if the laminar state is stable to infinitesimal amplitude perturbations \cite{larson92,graham2014, moro07}.  

Newtonian pipe flow is known to be linearly stable at all Reynolds numbers ($Re$) \cite{GILL65,trefethen03,drazin,schmid-henningson,kers05}. By carefully minimizing external perturbations, laminar flow has been maintained in experiments upto $Re \sim 100,000$; in contrast, when forced with finite amplitude disturbances, transition occurs around an $Re$ of 2000 \cite{pfenniger,mullin03,cohen07,mullin11,hof07}. Theoretically, this sub-critical scenario is explained by the appearance, above a threshold $Re$, of non-trivial three-dimensional solutions of the Navier-Stokes equations (termed exact coherent states) which are disconnected from the laminar state \cite{wall98,kers05,hof07}. Rectilinear shearing flows, including pipe flow, of dilute polymer solutions are also believed to be linearly stable at all Deborah numbers ($De$) in the inertialess limit $(Re=0)$ \cite{larson92,wilson99, renardy86, lenov67};  $De$ here being the ratio of the polymer relaxation time to the flow time scale. A non-linear mechanism has been proposed for transition to (elastic) turbulence in such flows, where an initial finite amplitude perturbation induces curved streamlines, which then become unstable to a hoop stress driven elastic instability that operates at linear order in canonical curvilinear geometries \cite{larson90,shaqfeh96,moro07,moro13,moro03}. Theoretical work explaining transition, and turbulent drag reduction, at finite $Re$ and $De$ has focused on the modification of the Newtonian scenario, by mapping the domain of existence of the exact coherent states in the $Re$-$De$ plane \cite{graham2014,graham02,graham04,larson06}. That these finite amplitude solutions do not exist above a critical $De$, for fixed $Re$, is indicative of a distinct transition mechanism at larger $De$ \cite{graham2014,graham06}. A separate line of work has focused on the linear transient growth of disturbances from a stable laminar state \cite{kumar08,kumar13, zaki14}. In summary, the viewpoint with regard to transition in dilute polymer solutions is rooted in the (assumed) linear stability of the laminar state everywhere in the $Re$-$De$ plane \citep{larson92,graham2014, moro07}. This is despite the absence of a rigorous linear stability analysis for pipe flow valid at large $Re$ and $De$.
   
There have, however, been scattered observations that point to a linear instability in pipe flow experiments involving dilute polymer solutions. In a series of experiments in the 1960s and 70s, transition to turbulence was observed in dilute polymer solutions, at Reynolds numbers much lower than the Newtonian threshold by several groups, the phenomenon being dubbed `early turbulence' \cite{little1974, little1972, little1970, forame1973, tamir1964, adrian69, hoyt77, zakin77, jones66, jones76}. Later, Draad et al. \cite{draad98} observed an order of magnitude reduction in the natural (unforced) transition $Re$ for a polymer solution. More recently, Samanta et al. \cite{MORO2013} studied transition in polyacrylamide solutions, in smaller diameter pipes, thereby accessing higher Deborah numbers. In a 4 mm diameter pipe, the transition process for concentrations  lesser than 200 ppm was analogous to the Newtonian one with forced and natural transitions occurring at disparate Reynolds numbers. In sharp contrast, for the 500 ppm solution, the transition occured at $Re \sim 800$ independent of the perturbation amplitude. Further, spatially localized structures (puffs), characteristic of the bistability associated with the Newtonian sub-critical transition \cite{wygnanski75,wygnanski73,barkley16}, were absent. Subsequently, this novel transitional pathway, connecting the laminar state to a novel elasto-inertial turbulent state, has been demonstrated over a much wider parameter range \cite{hof2017}. 

Although a linear instability has occasionally been speculated upon \cite{graham2014,little1972}, the (unstated) view in the field assumes otherwise \cite{moro07,MORO2013,dubief17}. Contrary to this view, we demonstrate in this letter that the laminar state is not linearly stable everywhere in the $Re$-$De$ plane, thereby pointing to a pathway to turbulence in viscoelastic pipe flow which has thus far remained unexplored.

The governing system of equations for an incompressible viscoelastic fluid (in non-dimensional form) is
\begin{equation}
Re(\frac{\partial}{\partial t} +  \bld u \bcdot \bnabla) \bld u=-\bnabla p+\frac{1-\beta}{De} \bnabla \bcdot \boldsymbol{A_p} +\beta \nabla^2 {\bld u}, \bnabla \bcdot {\bld u} =0
\label{prleq1},
\end{equation}
where $\bld u, p $ and $\boldsymbol{A_p}$ are the velocity field, pressure and the elastic stress tensor, respectively. The relevant non-dimensional parameters are $\beta = \frac{\mu_s}{\mu_p + \mu_s}$, $De = \frac{U_s \tau}{a}$ and $Re = \frac{\rho U_s a}{\mu_{s} + \mu_{p}}$  where $\mu_s$ and $\mu_p$ are the solvent and polymer contributions to the viscosity, $\tau$ the relaxation time of the polymer molecule, $\rho$ the density of the fluid, $a$ the pipe radius and $U_s$ (the centerline velocity) the imposed velocity scale (for steady laminar flow, the Deborah number $De$ is the same as the Weissenberg number $Wi$ \cite{tanner}). The elastic  stress is assumed to be governed by the Oldroyd-B constitutive equation, corresponding to polymer molecules in the solution being modeled as non-interacting Hookean dumbbells. This gives $\boldsymbol{A_p} \propto \langle \boldsymbol{RR} \rangle$, where $ \boldsymbol{R}$ is the dumbbell end-to-end vector and $\langle . \rangle$ denotes a configurational average. The affine deformation of $\boldsymbol{R}$, together with linear relaxation in a time $\tau$, leads to the following equation for $\boldsymbol{A_p}$  \cite{larson88}: 
\begin{equation}
(\frac{\partial}{\partial t} + \bld u \bcdot \bnabla) \boldsymbol{A_p} - \boldsymbol{A_p} \bcdot \bnabla \bld u - (\bnabla \bld u)^{\dagger} \bcdot \boldsymbol{A_p} =  - \frac{\boldsymbol{A_p}-\textbf{I}}{De}.
\label{prleq2}
\end{equation}

The Oldroyd-B model predicts a shear-rate independent viscosity and first normal stress coefficient in viscometric flows \cite{larson88}. It has been shown to reproduce observations of linear instabilities in polymer solutions in various curvilinear \cite{shaqfeh96} and extensional flows \cite{poole07} as well as the inertialess non-linear instability in rectilinear shearing flows \cite{moro03,moro07}, and is thus appropriate for a first effort. For $\beta=0$, (\ref{prleq1}) and (\ref{prleq2}) reduce to the Upper Convected Maxwell (UCM) model, with no solvent stress contribution.

\begin{figure}
\includegraphics[scale=0.85]{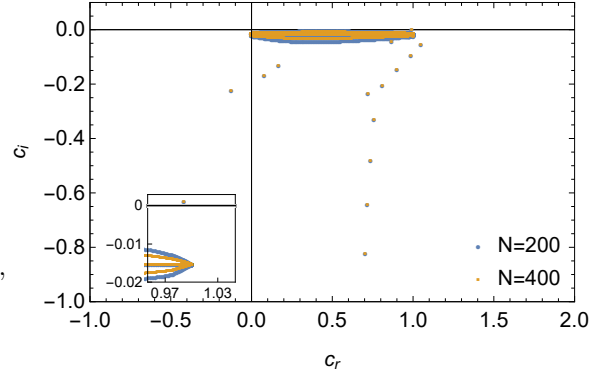}
\caption{\label{spectra}Eigenspectrum for pipe flow of an Oldroyd-B fluid for $Re=800, De = 65, \beta = 0.65$ and $k=1$ (for $N =200$ and $400)$; the inset zooms into the region around the unstable mode.}
\end{figure}

The laminar pipe flow profile for an Oldroyd-B fluid is the same as the Newtonian one, $U=1-r^2$. An associated first normal stress difference, $N_1=8 r^2 De^2$, arises owing to the polymer molecules being stretched and aligned with the flow. Assuming infinitesimal perturbations, $ \bld u =U+\boldsymbol{u^{\prime}}, \boldsymbol{A_p}=\boldsymbol{A}+\boldsymbol{a^{\prime}}, p=p_0+p^{\prime}$, of the normal mode form, $f^{\prime}=\hat{f}(r)e^{(ik(z-ct)+i m \theta)}$ (where $k$ and $m$ are the axial and azimuhtal wavenumbers, respectively), and linearizing about the aforementioned base-state, one obtains the following eigenvalue problem for pipe flow,
\begin{equation}
\boldsymbol{\mathcal{L}}\boldsymbol{\hat{f}}=c\boldsymbol{\hat{f}} \nonumber ,
\end{equation}
such that $c = c_r+i c_i \equiv c(Re,De,k,m,\beta)$ where $c_r$ is the wave speed and $c_i$ the gowth rate;  $c_i > 0 $ implies exponentially growing normal modes. We only consider axisymmetric perturbations $(m=0)$ in this letter, since non-axisymmetric disturbances were found to be stable over the parameter range considered. Two different numerical methods are used to solve the eigenvalue problem: a spectral collocation method in which the perturbation fields are expanded in terms of Chebyshev polynomials \cite{boyd} and a shooting method that numerically integrates the governing equations and iterates over the eigenvalue  $c$ (with a Newton-Raphson procedure) in order to satisfy the boundary conditions \cite{schmid-henningson}. We have verified our numerical schemes by reproducing earlier stability results for plane Poiseuille flow of an Oldroyd-B fluid \cite{sureshkumar95, zaki2013} and for Newtonian pipe flow \cite{schmid-henningson}. To avoid spurious modes, convergence was checked, for both eigenvalues and eigenfunctions, with respect to $N$ (the number of Chebyshev polynomials in the spectral expansion) as well as against the shooting method. The only prior work on linear stability of viscoelastic pipe flow neglected the convected derivative in (\ref{prleq2}) and hence is of restricted validity \cite{hansen1973}. 

\begin{figure}
\includegraphics[scale=0.625]{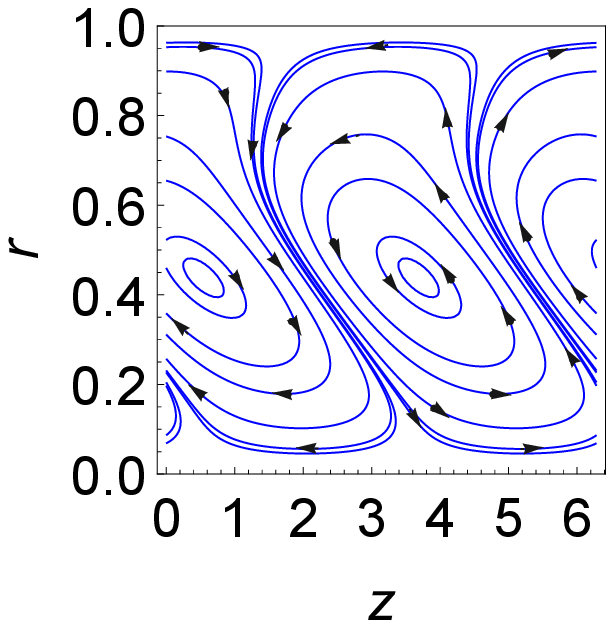}
\includegraphics[scale=0.625]{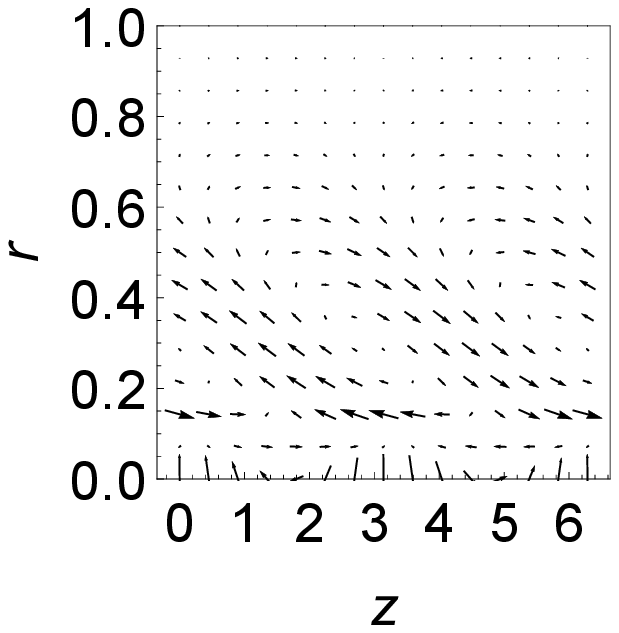}
\caption{Perturbation velocity (left) and polymer force (right) fields for the unstable mode for $Re=800, De = 65$, $\beta = 0.6$ and $k=1$.}
\label{vectorplot}
\end{figure}

The eigenvalue spectrum in figure \ref{spectra}, for $Re=800,De=65,\beta=0.65,k=1$, shows a single unstable mode, multiple damped discrete modes and a pair of continuous spectra (these appear as balloons due to the finite discretization). The continuous spectrum eigenvalues are given by $c=U-i/(k De)$ and $c=U-i/(\beta k De)$  and correspond to singular modes whose decay rates are set by the polymeric stress relaxation \citep{wilson99,KUP05,graham98,sureshkumar95,grillet02,zaki2013}. The unstable mode is an axisymmetric center-mode propagating at a speed close to the base-state maximum. Figure \ref{vectorplot} shows the associated perturbation velocity and polymer force density ($\bnabla \bcdot \boldsymbol{a^{\prime}}$) fields. The polymer force field is localized near the centerline and reinforces the velocity field, leading to the instability. The radial structure of the polymer forcing is reminiscent of recent simulations of elasto-inertial turbulence (for plane Poiseuille flow) wherein regions of high polymer stretch, localized in the gradient direction, were observed \cite{MORO2013, dubief13, dubief17}. 

\begin{figure}
\includegraphics[scale=0.425]{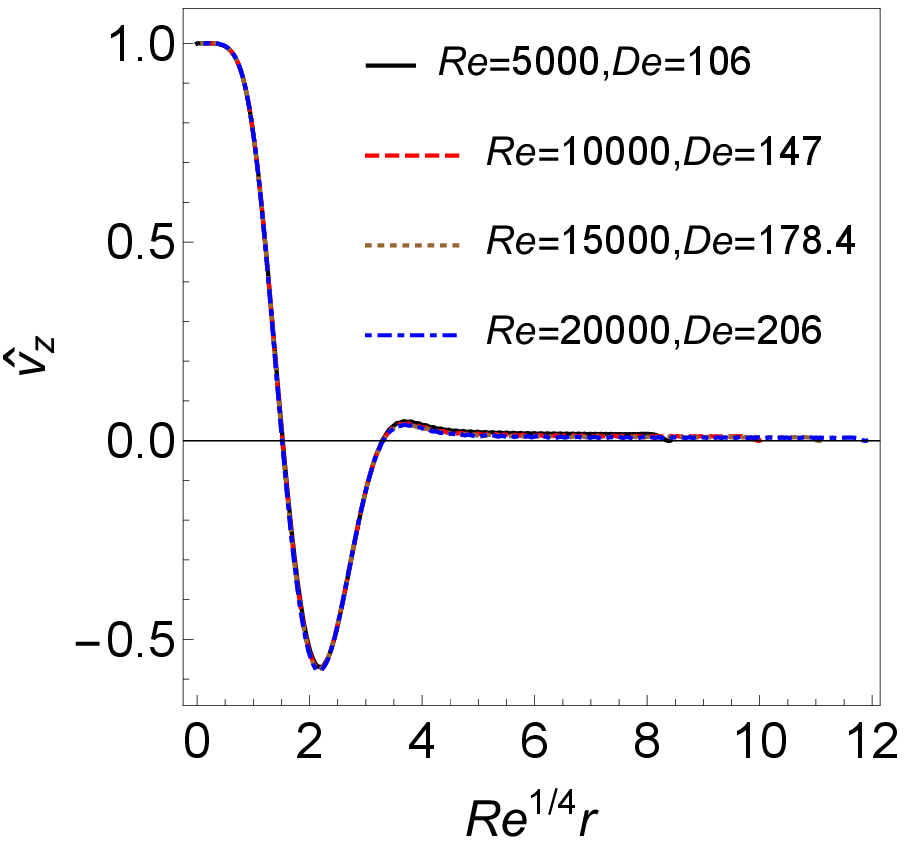}
\includegraphics[scale=0.425]{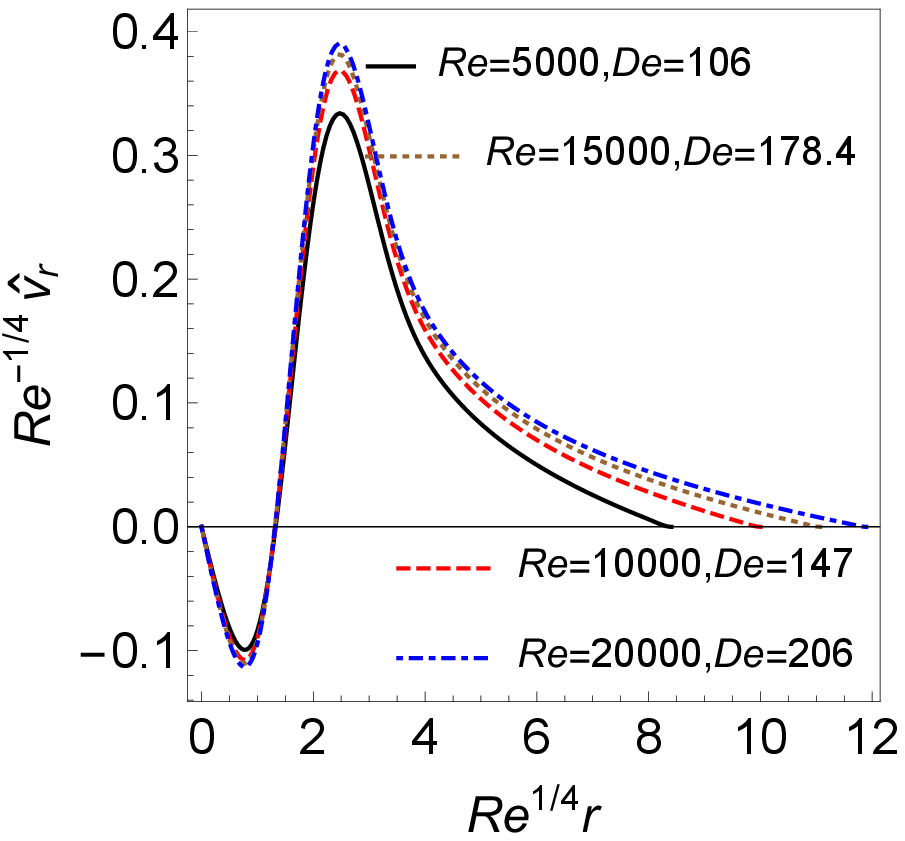}
\caption {The unstable center-mode eigenfunctions for the axial velocity (left) and radial velocity (right) in scaled boundary-layer coordinates in the limit $Re \rightarrow \infty$ and $De \rightarrow \infty$ for a fixed $De/Re^{1/2}$ ($k = 1$ and $\beta = 0.5$).}
\label{eigenfunctions}
\end{figure}

In the limit $Re,De \rightarrow \infty$ with $De/Re^{1/2}$ (and $k$) fixed, the unstable eigenfunctions become increasingly localized in a boundary layer of $\mathcal{O}(Re^{-1/4})$ around the centerline (figure \ref{eigenfunctions}). Viscous diffusion balances inertia in this boundary layer, analogous to a Newtonian center-mode  \cite{GILL65}, and for the perturbation polymeric stress to stay comparable requires $De \sim \mathcal{O}(Re^{1/2})$. The instability thus requires a balance of inertia, viscous and elastic polymer stresses close to the centerline. The centerline localization is in contrast to the original Newtonian and the elastically modified Tollmien-Schlichting instability for plane Poiseuille flow, where the eigenfunction is localized near the channel walls for large $Re$ \cite{schmid-henningson, zaki2013, sureshkumar95}.

\begin{figure}[!tbp]
  \centering
  \hfill
    \includegraphics[scale=0.4]{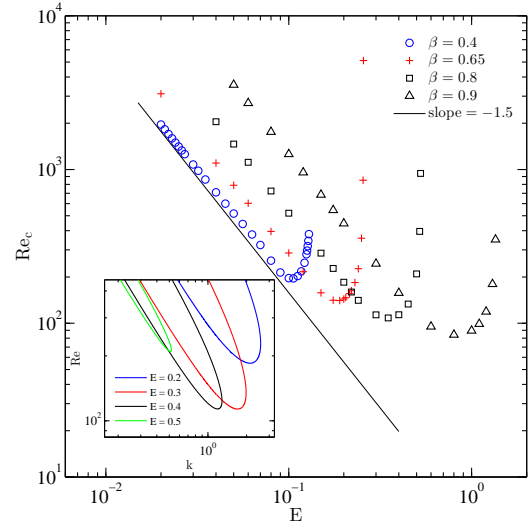}
\caption{The critical Reynolds number, $Re_c$, as a function of $E$ for different viscosity ratios, $\beta$. The inset shows neutral curves in the $Re$-$k$ plane for $\beta = 0.8$ for different $E$.}
\label{region1}
\end{figure}

For a given polymer solution, the elasticity number $E = \frac{De}{Re}$ and $\beta$ are fixed, and independent of the imposed flow velocity. Hence, in figure \ref{region1}, we characterize the instability in terms of a critical Reynolds number, $Re_c$, as a function of $E$ and $\beta$. At a given $E$ and $\beta$, $Re_c$ is found by minimizing the threshold $Re$ over all $k$. Both branches of the neutral curve in the $Re$-$k$ plane (see figure \ref{region1} inset) show the expected long wavelength scaling, $Re \sim \mathcal{O}(1/k)$, for $k \rightarrow 0$. Further, figure \ref{region1} shows that, at a fixed $\beta$, the left branch of the $Re_c$-$E$ curve is such that $Re_c \propto E^{-3/2}$ with $k_c \propto E^{-1/2}$ (not shown). The right branch is almost vertical and hence for a given $\beta$ there appears to be an $E_{crit}$ such that the instability does not exist for any $Re$ for $E>E_{crit}$; while for $E < E_{crit}$, the laminar state is always unstable at large enough $Re$. As $\beta$ is increased, the minima in the $Re_c$-$E$ curves shift to higher $E$ and lower $Re_c$ and the unstable region in the $Re_c$-$E$ plane increases in extent.  

\begin{figure}
\includegraphics[scale=0.5]{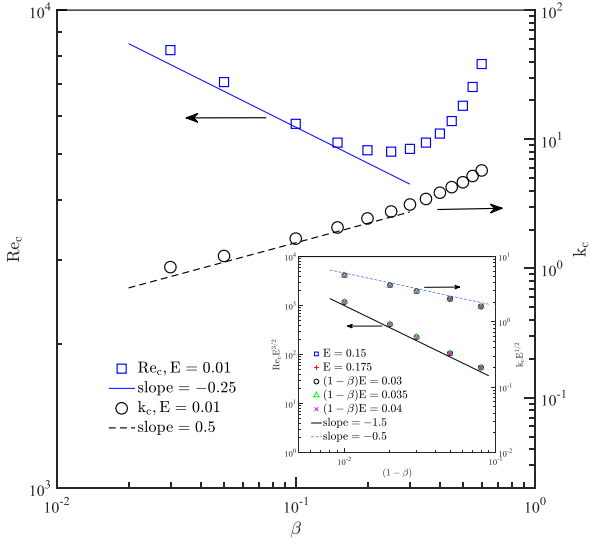}
\caption{The critical Reynolds number, $Re_c$, as a function of the viscosity ratio $\beta$ for $E = 0.01$. The inset shows the scaling behavior in the dual limit $E \rightarrow 0, \beta \rightarrow 1$.}
\label{region2}
\end{figure}

Figure~\ref{region2} shows the expected absence of the instability in the Newtonian limit and its surprising absence in the UCM limit. In the Newtonian limit ($\beta \rightarrow 1$), $Re_c \propto (1-\beta)^{-3/2}$ and $k_c \propto (1-\beta)^{-1/2}$. The $\beta$ and $E$ scalings above may be combined in the dual limit $E(1-\beta) \ll 1$ and $\beta \rightarrow 1$, so that $Re_c \propto (E (1-\beta))^{-3/2}$ and $k_c \propto (E (1-\beta))^{-1/2}$ (figure \ref{region2} inset). Thus, the instability survives provided $E \propto (1-\beta)^{-1}$, which ensures that the perturbation polymer stress remains of order unity. For a given $Re$ this implies that the minimum $De$ for which the instability exists diverges in the Newtonian limit. In all approaches to the Newtonian limit, the axial wavelength ($k_c^{-1}$) of the center-mode becomes comparable to the centerline boundary layer thickness ($O(\epsilon^{1/2})$, $\epsilon$ being the relevant small parameter). For $k_c \gg 1$, $Re_c \propto k_c^{3}$ ensures a balance between inertia and viscous stresses and, alongwith $k_c \propto \epsilon^{-1/2}$, predicts $Re_c \propto \epsilon^{-3/2}$; consistent with the observed scalings. 

In contrast to figure \ref{region2} which shows an asymptotic scaling for $Re_c$, for $\beta \rightarrow 1$, an analysis of the spectrum for fixed $Re$, $E$ and $k$, shows that the instability doesn't persist until the Newtonian limit. Instead, the unstable center-mode becomes stable at a finite $(1-\beta)$, eventually falling off into the continuous spectrum with the polymer force field becoming singular. In the UCM limit ($\beta \rightarrow 0$), $Re_c$ is shown to diverge as $\beta^{-1/4}$ in figure \ref{region2}, for $E=0.01$, but the associated critical wavenumber ($k_c$) decreases as $\beta^{1/2}$ for small $\beta$. The absence of the instability for $\beta =0$ reinforces the idea that all three physical effects (inertia, the viscous solvent stress and the elastic polymeric stress) are essential for the center-mode instability. This is in contrast to the expectation that the solvent stress generally plays a stabilizing role in elastic instabilities \cite{shaqfeh96, zaki2013}.

The instability is predicted to exist over a wide range of $Re$. The regime $Re_c \sim \mathcal{O}(100)$, $E \sim \mathcal{O}(1)$ is achievable in microfluidic devices \cite{kumaran2017,tamir1964}. For $(E,\beta)$ values such that $Re_c$ is $\mathcal{O}(2000)$ or greater, pertinent to macroscopic geometries, the sub-critical Newtonian transition might mask the linear instability unless external perturbations are carefully minimized. A natural transition $Re$ of around $8000$ was reported by Draad et al. \cite{draad98} for a 20 ppm solution of partially hydrolysed polyacrylamide in demineralized water ($\beta \sim 0.1$ based on the zero shear viscosity); as opposed to the much higher transition $Re$ of $60,000$ for Newtonian fluids for their experimental facility. Our calculations do yield an unstable mode at the corresponding $Re$ and $\beta$, for $E = 0.01$, although the strong shear thinning exhibited by the solution prevents a quantitative comparison. Similar observations of a significantly lower natural transition $Re$ have been reported for dilute solutions of polyethylene oxide \cite{abernathy72}. The instability also qualitatively explains the observations of `early turbulence' in \cite{little1974, little1972, little1970, tamir1964, adrian69, jones66}. For the $500$ ppm polyacrylamide solution used by Samanta et al. \cite{MORO2013}, $\beta = 0.65$. The $Re_c$ for the instability at this $\beta$ is well below that at which Newtonian turbulence sets in. This is in qualitative agreement with the experiment where transition was reported at $Re \sim 800$. However, the minimum $E$ required for the center-mode instability is around $0.05$, which is an order of magnitude larger than the experimentally reported value of $0.004$, based on a measurement of the relaxation time using a capillary break-up elongational rheometer (CaBER). This discrepancy may be attributed to the known difficulty in associating the time inferred from CaBER measurements to the relaxation rate relevant to the Oldroyd-B model \cite{samanta13,mckinley06,kumaran2017,wagner10,shankar13}. The ($Re$-$De$-$\beta$) dependent threshold of the center-mode instability calls for a re-examination of the expectation that early transition, even in the absence of finite amplitude perturbations, is governed by a critical $De$, regardless of $Re$ \cite{tamir1964,MORO2013}. We hope that the first theoretical evidence for the laminar state being unstable provided by this letter would motivate the search for a decisive experimental demonstration. 


Observations of pressure-driven flow through a channel of a polyacrylamide solution becoming turbulent at $Re \sim 350$, $De \sim 250$ and $\beta = 0.92$ were reported in \cite{kumaran2017}. We have verified that a center-mode instability, similar to the one described above for pipe flow, exists at these parameter values for plane Poiseuille flow of an Oldroyd-B fluid; the details will be reported elsewhere \cite{ganesh18}. Plane Couette flow was, however, found to be stable at all $Re$ and $De$ values examined. Since the polymers are only weakly stretched near the centerline, the center-mode nature of the instability suggests the relative unimportance of finite extensibility of the polymer chains. Indeed, preliminary work using the FENE-P constitutive equation, typically employed in simulation studies of viscoelastic turbulence \cite{dubief13,sureshkumar1997direct,dubief2004coherent}, shows that the instability persists in the presence of shear thinning \cite{ganesh18}.

The instability described in this letter should form the first step in describing a new pathway to turbulence, and possibly the maximum drag reduction state, in dilute polymer solutions. The general mechanism will be applicable to inertial flows of other viscoelastic fluids such as wormlike micellar surfactant solutions which show drag reduction \cite{zakin98,samanta13}. At the linear instability threshold, novel elasto-inertial traveling wave solutions, associated with the unstable center-mode eigenfunctions, would be created in a Hopf bifurcation from the laminar state \cite{barkley90,soibelman91}. The implied contrast between the state space for viscoelastic pipe flow and the Newtonian one will have fundamental consequences for the dynamical systems interpretation of the maximum drag reduction state which, currently, crucially relies on a similarity between the two \cite{larson06,graham2014,graham02,dubief17,xi2012}. The aforementioned traveling wave solutions and associated phase space structures could also be relevant for describing two-dimensional elasto-inertial turbulence, recently observed in simulations \cite{dubief17,dubief13}. Practically, a detailed understanding of the transitional pathway associated with the instability would help develop control strategies to induce early (or delayed) transition to turbulence, which would be of special relevance to microfluidic devices \cite{squires05,li12,kumaran15,hong16,lim14}.     

\emph{} P.G. and I.C. contributed equally to this work. 

\bibliography{refer}

\end{document}